\documentclass[12pt]{article}

\begin{document}

\center{\large The Relationship between Bare and Renormalized Couplings in Scalar Electrodynamics}

\center{D.G.C. McKeon\\Department of Applied Mathematics\\University of Western Ontario\\London Ontario Canada\\N6A 5B7}

\abstract{When using dimensional regularization, the bare couplings are expressed as a power series in $(2-n/2)^{-1}$ where $n$ is the number of dimensions. It is shown how the renormalization group can be used to sum the leading pole, next to leading pole etc. contributions to these sums in scalar electrodynamics (or any theory with multiple couplings.)}

\vskip 1cm
\baselineskip 18pt

   In scalar electrodynamics, the two coupling constants are the quartic scalar coupling $\lambda$ and the square of the electric charge $\alpha=e^{2}$. These renormalized couplings are related to their bare counterparts by the equations 

\begin{equation}
\lambda_{B}=\mu^{2\epsilon}\lambda\sum^{\infty}_{\nu=0}\frac{a^{(\lambda)}_{\nu}(\lambda,\alpha)}{\epsilon^{\nu}}
\end{equation}
\begin{equation}
\alpha_{B}=\mu^{2\epsilon}\alpha\sum^{\infty}_{\nu=0}\frac{a^{(\alpha)}_{\nu}(\lambda,\alpha)}{\epsilon^{\nu}}
\end{equation}
where $\epsilon=2-n/2$ and $\mu^{2}$ is a dimensionful parameter needed to ensure that $\lambda$ and $\alpha$ are dimensionless in $n$ dimensions \cite{1}. These functions $a^{(\lambda)}_{\nu}$ and $a^{(\alpha)}_{\nu}$ can themselves be expanded  

\begin{equation}
a^{(g)}_{\nu}(\lambda,\alpha)=\sum^{\infty}_{n=\nu}\sum^{n}_{m=0}a^{(g)\nu}_{n-m,m}\lambda^{n-m}\alpha^{m}
\end{equation}
where $g$ is either $\lambda$ or $\alpha$.
   
   As the bare couplings are independent of $\mu^{2}$, variations of $\mu^{2}$ are to be compensated by changes in $\lambda$ and $\alpha$ so that the renormalization group (RG) equations
\begin{equation}
\left(\mu^{2}\frac{\partial}{\partial\mu^{2}}+\mu^{2}\frac{\partial\lambda}{\partial\mu^{2}}+\mu^{2}\frac{\partial\alpha}{\partial\mu^{2}}\right)g_{B}=0
\end{equation}
are satisfied.
   The RG functions are 
\begin{equation}
\tilde{\beta}^{(g)}(\lambda,\alpha)=\mu^{2}\frac{\partial g}{\partial\mu^{2}}=-\epsilon g +\beta^{(g)}(\lambda,\alpha)
\end{equation}
in the renormalization scheme in which
\begin{equation}
a^{(g)}_{0}(\lambda,\alpha)=1;
\end{equation}
this is the "minimal subtraction" (MS) scheme.
   The functions $\beta^{(\lambda)}$ and $\beta^{(\alpha)}$ have the expansions 
\begin{equation}
\beta^{(g)}=\sum^{\infty}_{n=2}\beta^{(g)}_{n}
\end{equation}
where
\begin{equation}
\beta^{(g)}_{n}=\sum^{n}_{m=0}b^{(g)}_{n-m,m}\lambda^{n-m}\alpha^{m}
\end{equation}

   A discussion of the relationship between the bare and renormalized couplings in the simpler case in which there is a single coupling appears in \cite{2,3}. We now extend this analysis to scalar electrodynamics using methods similar to those employed in the effective potential \cite{4} and the kinetic term \cite{5} in the effective action.

    The first step is to define 
\begin{equation}
p^{(g)l}_{n+l}=\sum^{n+l}_{m=0}a^{(g)}_{n+l-m,m}\lambda^{n+l-m}\alpha^{m}
\end{equation}    
so that by eqs.(1,3)     
\begin{equation}
g_{B}=\mu^{2\epsilon}g\sum^{\infty}_{n=0}\sum^{\infty}_{l=0}p^{(g)l}_{n+l} \epsilon^{-l}.    
\end{equation}    
Using eqs.(9,10), eq.(4) becomes
\begin{eqnarray}
&&\left[\mu^{2}\frac{\partial}{\partial\mu^{2}}+(-\epsilon\lambda+\sum^{\infty}_{n=2}\beta^{(\lambda)}_{n})\frac{\partial}{\partial\lambda}
 +(-\epsilon\alpha 
+\sum^{\infty}_{n=2}\beta^{(\alpha)}_{n})\frac{\partial}{\partial\alpha}\right]\nonumber\\
&&\left[g\mu^{2\epsilon}\sum^{\infty}_{l=0}\sum^{\infty}_{n=0}p^{(g)l}_{n+l}\epsilon^{-l}\right]=0.
\end{eqnarray}
The term in eq.(10) of order $\epsilon$ is automatically satisfied if $p^{(g)0}_{n}=\delta_{n,0}$. Specializing to the case $g=\lambda$ ($g=\alpha$ can be treated the same way)we find that eq.(10) reduces to 
\begin{eqnarray}
&& \sum^{\infty}_{l=0}\sum^{\infty}_{n=0}\epsilon^{-l} \left[-\lambda(\lambda\frac{\partial}{\partial\lambda}+\alpha\frac{\partial}{\partial\alpha})p^{(\lambda)l+1}_{n+l+1} \right. \nonumber \\
&& \left.  +\sum^{\infty}_{p=2}\left(\beta^{(\lambda)}_{p}(1+\lambda\frac{\partial}{\partial\lambda})+\beta^{\alpha}_{p}\lambda\frac{\partial}{\partial\alpha}\right)p^{(\lambda)l}_{n+l}\right]=0.
\end{eqnarray}
Upon defining
\begin{equation}
S^{(\lambda)}_{n}=\sum^{\infty}_{l=0}p^{(\lambda)l}_{n+l}\epsilon^{-l}
\end{equation}
then the sum of the 'leading poles' (LP) is given by $S^{(\lambda)}_{0}$, the 'next to leading poles' (NLP) is given by $S^{(\lambda)}_{1}$ etc.

   If now we define $\xi\equiv\lambda/\epsilon$, $\zeta\equiv\alpha/\epsilon$, then by eqs.(11,12,13) it follows that if eq.(12) is satisfied at each order in the couplings then 
\begin{eqnarray}
& -& \xi \left( \xi\frac{\partial}{\partial\xi} + \zeta\frac{\partial}{\partial\zeta}\right) S^{(\lambda)}_{n}(\xi,\zeta) \nonumber\\
&+&\sum^{n}_{k=0}\left(\beta^{(\lambda)}_{n+2-k} \left(1+\xi\frac{\partial}{\partial\xi} \right)+\beta^{(\alpha)}_{n+2-k}\xi\frac{\partial}{\partial\zeta}\right)S^{(\lambda)}_{k}(\xi,\zeta)=0.
\end{eqnarray}

For $n=0$, eq.(14) becomes 
\begin{equation}
\left[(\beta^{(\lambda)}_{2}\xi-\xi^{2})\frac{\partial}{\partial\xi}+(\beta^{(\alpha)}_{2}\xi-\xi\zeta)\frac{\partial}{\partial\zeta}+\beta^{\lambda}_{2}\right]S^{\lambda}_{0}=0.
\end{equation}
This equation can be solved using the method of characteristics \cite{4,5}. We begin by rewritting eq.(15) as
\begin{equation}
\left[\frac{\partial}{\partial\zeta}+\frac{\beta^{(\lambda)}_{2}-\xi}{\beta^{(\alpha)}_{2}-\zeta}\frac{\partial}{\partial\xi}+\frac{\beta^{(\lambda)}_{2}}{\beta^{(\alpha}_{2}\xi-\xi\zeta}\right]S^{(\lambda)}_{0}(\xi,\zeta)=0 
\end{equation}
and then introducing the auxiliary function $\bar{\xi}(\zeta)$ so that
\begin{equation}
\frac{d\bar{\xi}(\zeta)}{d\zeta}=\frac{\beta^{(\lambda)}_{2}(\bar{\xi}(\zeta),\zeta)-\bar{\xi}(\zeta)}{\beta^{(\alpha)}_{2}(\bar{\xi}(\zeta),\zeta)-\zeta}
\end{equation}
with $\bar{\xi}(0)=\xi$. Eq. (16) then becomes
\begin{equation}
\left[\frac{d}{d\zeta}+ \frac{\beta^{(\lambda)}_{2}(\bar{\xi}(\zeta),\zeta)}{[\beta^{(\alpha)}_2(\bar{\xi}(\zeta),\zeta)-\zeta]\bar{\xi}(\zeta)}\right]S^{(\lambda)}_{0}(\bar{\xi}(\zeta),\zeta)=0.
\end{equation}
with $\bar{\xi}(0)=\xi$.

  Eq. (18) has a solution
\begin{equation}
S^{(\lambda)}_{0}=exp\left(-\int^{\zeta}_{0}d\eta\frac{\beta^{(\lambda)}_{2}(\bar{\xi}(\eta),\eta)}{[\beta^{(\alpha)}_{2}(\bar{\xi}(\eta),\eta)-\eta]\bar{\xi}(\eta)}\right)S^{(\lambda)}_{0}(\bar{\xi}(0),0).
\end{equation}
For $S^{(\lambda)}_{0}(\bar{\xi}(0),0)=S^{(\lambda)}_{0}(\xi,0)\equiv\sigma_{0}(\xi)$, we impose the condition that this boundary term be given by the LP term in a scalar theory with a complex scalar field. (That is, we 'turn off' the electric charge in this limiting case.)

  In this simpler model \cite{1}, we have
\begin{equation}
\mu^{2}\frac{\partial\lambda}{\partial\mu^{2}}=\tilde{\beta}(\lambda)=-\epsilon\lambda+\sum^{\infty}_{n=2}b_{n}\lambda^{n}\equiv-\epsilon\lambda+\beta(\lambda)
\end{equation}
and
\begin{equation}
\left(\mu^{2}\frac{\partial}{\partial\mu^{2}}+\tilde{\beta}(\lambda)\frac{\partial}{\partial\lambda}\right)\lambda_{B} =0
\end{equation}
with
\begin{equation}
\lambda_{B}=\mu^{2\epsilon}\lambda\sum^{\infty}_{n=0}\sum^{n}_{m=0}a_{n-m,m}\lambda^{n-m}\epsilon^{-m}=\mu^{2\epsilon}\lambda\sum^{\infty}_{n=0}\lambda^{n}\sigma_{n}(\xi)
\end{equation} 
 
where $a_{n,0}=\delta_{n,0}$ and $\sigma_{n}(\xi)\equiv\sum^{\infty}_{m=0}a_{n+m,m}(\lambda/\epsilon)^{m}$. Together, eqs. (20-22) show that for $n=1,2,...$
\begin{equation}
(1-n)\sigma_{n-1}-\xi\sigma^{'}_{n-1}+\sum^{n+1}_{k=2}\xi b_{k}\left[(n+2-k)\sigma_{n+1-k}+\xi\sigma^{'}_{n+1-k}\right]=0,
\end{equation}
so that if $n=1$
\begin{equation}
-(1-b_{2}\xi)\sigma^{'}_{0}+b_{2}\sigma_{0}=0.
\end{equation}

   When using MS $\sigma_{n}(0)=\delta_{n,0}$; with this boundary condition, eq. (24) has the solution 
\begin{equation}
\sigma_{0}(\xi)=(1-b_{2}\xi)^{-1}
\end{equation}
which we take to be the function $S^{(\lambda)}_{0}(\bar{\xi}(0),0)$ in eq. (19).

   Similarly, one can find $S^{(\alpha)}_{0}(\xi,\eta)$.

   When n=1, eq. (14) yields
\begin{eqnarray}
&-&\xi\left(\xi\frac{\partial}{\partial\xi}+\zeta\frac{\partial}{\partial\zeta}\right)S^{(\lambda)}_{1} + \left(\beta^{(\lambda)}_{3}(1+\xi\frac{\partial}{\partial\xi})+\beta^{(\alpha)}_{3}\xi\frac{\partial}{\partial\zeta}\right)S^{(\lambda)}_{0}
\nonumber\\
&+&\left(\beta^{(\lambda)}_{2}(1+\xi\frac{\partial}{\partial\xi})+\beta^{(\alpha)}_{2}\xi\frac{\partial}{\partial\zeta}\right)S^{(\lambda)}_{1}=0.
\end{eqnarray}

   Again introducing the auxiliary function $\bar{\xi}(\zeta)$ of eq. (17), eq. (26) becomes
\begin{eqnarray}
&&\left(\frac{d}{d\zeta}+\frac{\beta^{(\lambda)}_{2}}{(\beta^{(\alpha)}_{2}-\zeta)\bar{\xi}}\right)S^{(\lambda)}_{1}\nonumber \\
&+&\frac{1}{(\beta^{(\alpha)}_{2}-\zeta)\bar{\xi}}\left(\beta^{(\lambda)}_{3}(1+\bar{\xi}\frac{\partial}{\partial\bar{\xi}})+\beta^{(\alpha)}_{3}\bar{\xi}\frac{\partial}{\partial\zeta}\right)S^{(\lambda)}_{0}=0.
\end{eqnarray}       
    
   This equation has the solution
\begin{equation}
S^{\lambda}_{1}(\bar{\xi}(\eta),\eta)=exp\left(-\int^{\eta}_{0}d\tau p(\tau)\right)\left[\int^{\eta}_{0}e^{\int^{\tau}_{0}d\tau^{'}p(\tau^{'})}q(\tau)d\tau+\sigma_{1}(\xi)\right]    
\end{equation}    
where 
\begin{equation}
p(\tau)=\beta^{(\lambda)}_{2}(\bar{\xi}(\tau),\tau)\left[(\beta^{(\alpha)}_{2}(\bar{\xi}(\tau),\tau)-\tau)\bar{\xi}(\tau)\right]^{-1}
\end{equation}
and 
\begin{equation}
q(\tau)=\frac{-1}{(\beta^{(\alpha)}_{2}-\tau)\bar{\xi}}\left[\beta^{(\lambda)}_{3}(1+\bar{\xi}\frac{\partial}{\partial\bar{\xi}})+\beta^{(\alpha)}_{3}\bar{\xi}\frac{\partial}{\partial\tau}\right]S^{(\lambda)}_{0}(\bar{\xi}(\tau),\tau)
\end{equation}
when we have the boundary condition $S^{(\lambda)}_{1}(\bar{\xi}(0),0)=\sigma_{1}(\xi)$.     
    
   From eq. (23), it is possible to show that when $n=2$    
\begin{equation}
-\sigma_{1}-\xi\sigma^{'}_{1}+\xi b_{2}(2\sigma_{1}+\xi\sigma^{'}_{1})+\xi b_{3}(\sigma_{0}+\xi\sigma^{'}_{0})=0,    
\end{equation}    
so that
\begin{equation}
\sigma_{1}(\xi)=\frac{-b_{3}}{b^{2}_{2}}\left[\frac{1}{\xi(1-b_{2}\xi)}\right]\left[ln(1-b_{2}\xi)+\frac{1}{1-b_{2}\xi}-1\right].
\end{equation}    
 
   So also, $S^{(\lambda)}_{n}(\eta,\zeta)$ can be computed in terms of $\beta^{(\lambda)}_{2}..\beta^{(\lambda)}_{n+2}, \beta^{(\alpha)}_{2}..\beta^{(\alpha)}_{n+2}$ and $\sigma_{m}(\xi)$ for $n>2$. It is easily established by substitution of eqs. (1,2) into eq. (4) that $\beta^{(g)}$ itself is fixed by $a^{(g)}_{1}$.    

    In \cite{2,3} it is established that $\sigma_{n}\rightarrow 0$ as $\xi\rightarrow\infty$ (i.e. $\epsilon\rightarrow 0$). Since $S^{(\lambda)}_{n}$ is linear in $\sigma_{0}...\sigma_{n}$, we see that the bare couplings will themselves vanish in our model when the regulating parameter $\epsilon \rightarrow 0$, even though each term in the expansions of eqs. (1,2) diverges in this limit.
\begin{center}
\Large{Acknowledgements}
\end{center}
The author would like to thank Prof. G. Davies of Algoma University for his hospitality when this work was done. G. Reid was most helpful and R. Macleod had a useful suggestion with one of the equations.


\begin{thebibliography}{99}
\bibitem{1}G.'t Hooft, Nucl.Phys.B61, 455 (1973).
\bibitem{2}V.Elias and D.G.C.McKeon, Int.J.Mod.Phys.A18, 2395 (2003).
\bibitem{3}V.Elias, D.G.C.McKeon and T.N.Sherry, Int.J.Mod.Phys.A20, 1065 (2005).   
\bibitem{4}F.A.Chishtie, T.Hanif, D.G.C.McKeon and T.G.Steele, Phys.Rev.D77, 065007 (2008). 
\bibitem{5}F.A.Chishtie, J.Jia and D.G.C.McKeon, Phys.Rev.D76, 105006 (2007).
\end{thebibliography}
\end{document}